\documentclass[a4paper,12pt,english,german]{article}

\usepackage{caption}
\usepackage{subcaption}
\usepackage{mathtools}
\usepackage{amssymb,amsmath,scalefnt,cite}
\usepackage{titlesec}
\usepackage{graphicx}
\usepackage{multirow}
\usepackage{array}
\usepackage{hyperref}
\begin{document}

\centerline{\bf Invertibility as a witness of Markovianity of }
\centerline{\bf the quantum dynamical maps}

\bigskip

\centerline{  Jasmina Jekni\' c-Dugi\' c$^a$, Momir Arsenijevi\' c$^b$, Miroljub Dugi\' c$^{\ast b}$}

\bigskip

$^a$University of Ni\v s, Faculty of Science and Mathematics, Vi\v
segradska 33, 18000 Ni\v s, Serbia

$^b$University of Kragujevac, Faculty of Science, Radoja
Domanovi\' ca 12, 34000 Kragujevac, Serbia

{\bf Abstract}
 Markovianity of the quantum open system processes is a topic of the considerable current interest.
Typically,  invertibility   is assumed to be non-essential for Markovianity of the open-quantum-system dynamical maps.
Nevertheless, in this paper we  distinguish a class of physically important dynamical maps (processes) for which invertibility
is a necessary condition for Markovianity. Since every quantum-state tomography  directly provides information on invertibility
of the map, no optimization procedure is necessary for determining non-Markovianity regarding the considered class of dynamical processes.
On this basis we are able to provide a systematic insight and to distinguish mutual relations of the various approaches to quantum Markovianity.
Notably, for the processes out of the considered class of dynamical maps, various relations are allowed between divisibility, invertibility
and Markovianity of the dynamical maps.


Pacs: 03.65.Yz, 03.65.Db, 03.65.-w

\section{Introduction}\label{sec:1}

Realistic quantum systems are {\it open}, that is, in unavoidable contact with their surrounding environments \cite{breuer,rivas}.
Under assumption that an open system $S$ in contact with its environment $E$ constitutes the unitary whole $S+E$ that is subject of the unitary dynamics $\hat U(t,t_{\circ})$,
the open system's dynamics is determined by the time-dependent reduced density matrix $\hat\rho_S(t) = tr_E \left(\hat U(t,t_{\circ})\hat\rho_{SE}(t_{\circ})\hat U^{\dag}(t,t_{\circ}) \right)$
with the total $S+E$ system's state $\hat \rho_{SE}(t_{\circ})$ and the initial time instant $t_{\circ}$. In general,
finding the open system's state $\hat\rho_S(t)$ is  a highly non-trivial problem which can only be analytically solved in very
special scenarios and models.

 As a topic of special interest and a vivid ongoing research appear the so-called Markovian (as opposite to non-Markovian) dynamical processes (dynamical maps). Historically, the origin of interest in Markovian quantum dynamics is at least two-fold. On the one hand, the classical Markovian processes as a mathematical model are well investigated and extensively used. On the other hand, the
 celebrated Gorini-Kossakowski-Sudarshan-Lindblad (GKSL) form of the open quantum system's master equation \cite{gorini} is practically generally considered to be Markovian \cite{breuer,rivas,rivas1,breuer1,spanci,wiseman,milz}. From the practical point of view, presence of some kind of memory in the open system's dynamics may be dangerous or even fatal for certain purposes and applications, e.g. in the nascent field of quantum technology, notably in quantum information processing and quantum computation \cite{nielsen}, and quantum metrology \cite{metrol}.

 Unfortunately, it is notoriously hard to formulate a proper quantum-mechanical counterpart of the classical concept of Markovianity of dynamical maps so much so that relations between the different definitions and the related witnesses of Markovianity are still subject of vivid debate \cite{rivas1,breuer1,spanci,wiseman,milz}.

 Possibly the best known criteria for Markovianity that underpin virtually all the Markovian witnesses in the literature are the so-called CP-divisibility of dynamical maps (also known as the RHP criterion)
  \cite{rivas,rivasprl,rivas1} and the ''no information backflow'' (NIB) criterion (also known as the BLP criterion) \cite{breuer,breuer1,backflow}. For a dynamical map $\Phi(t,t_{\circ})$, CP-divisibility assumes existence of an intermediate map for non-zero initial instant of time, $\Phi(t,s), s>t_{\circ}$, which is completely positive (CP). The BLP criterion requires absence of information flow from the environment to the open system of interest. The two criteria are known not to be mutually equivalent \cite{rivas1,breuer1}: while CP-divisibility implies the BLP, the reverse is, in general, not true.

The map divisibility, i.e. existence of $\Phi(t,s),s>t_{\circ}$, is typically considered as a characteristic trait  while invertibility is regarded nonessential for Markovianity. Since existence of  $\Phi(t,s)$ is well understood for the invertible maps, i.e. for the maps that admit existence of the inverse, $\Phi^{-1}(t,t_{\circ})$,
there is ongoing research  devoted to ''going beyond invertibility'', e.g. \cite{hrusc1,datta,hrusc,anders,hall1}. The task is to introduce a proper concept of Markovian dynamics for the maps {\it not}
admitting the equality $\Phi(t,s)=\Phi(t,t_{\circ})\Phi^{-1}(s,t_{\circ})$. That is, the task is to provide Markovianity without resorting divisibility to invertibility of a map. Numerous dynamical models and scenarios provided
basically the \textit{model-dependent} relations regarding divisibility, invertibility and time locality of the dynamical processes. The general analyses of the role of invertibility and divisibility for existence of time-local
master equations \cite{anders,hall} are not conclusive. In Ref. \cite{anders}, the authors conjecture ''if it
can be guaranteed that the Hamiltonian does not have such
'artificial' features, then the time evolution is, in principle,
uniquely determined by a time-local master equation even in
cases when the time evolution is not invertible.''

In this paper we report on a significant role of invertibility for time-locality and therefore Markovianity of a process.
Our considerations provide a {\it systematic} and rather simple insight into the role of invertibility
(and divisibility) for the time-locality (and hence of Markovianity) of the process.

Our starting point is a {\it common point of agreement} of some important criteria of Markovianity.
Assume that a dynamical process admits a master equation (differential) form for the open system's reduced state $\hat\rho_S(t)$. This assumption is not really restrictive. As we demonstrate below
(cf. Section \hyperref[sec:3.2]{\ref{sec:3.2}}), non-differentiable processes can be straightforwardly described.
Then the mentioned common point is the requirement that in order for the process to be Markov, it  {\it must} be {\it local in time}:
\begin{equation}\label{eqn:1}
{d\hat\rho_S(t)\over dt} = \mathcal{L}_t\hat\rho_S(t).\tag{1.1}
\end{equation}

\noindent That is, time-locality of the generator $\mathcal{L}_t$ (the so called Liouvillian, which is a traceless linear operator on the Banach space of density matrices) is a {\it necessary condition} for a
process to be regarded Markovian. In other words: a (time-)differentiable dynamical process not admitting a master equation of the general form of eq.(\ref{eqn:1}) is necessarily {\it non-Markovian}.
It should be clearly stated: equation (\ref{eqn:1}) applies also to some non-Markovian processes. That is, time-locality of a master equation is necessary but not sufficient for Markovianity of the process.

For a \textit{physically important} class of differentiable processes, denoted $\mathcal{C}$, we prove that invertibility of a dynamical map is  {\it necessary and sufficient} condition for time locality of the related master equation.
Consequently, invertibility of a dynamical map in the class $\mathcal{C}$ is a necessary condition for Markovianity of the process. That is, a dynamical map $\Phi(t,t_{\circ})$ in the class $\mathcal{C}$ of dynamical
processes that is not invertible, i.e. does not admit existence of the inverse $\Phi^{-1}(t,t_{\circ})$, is {\it necessarily non-Markovian}.

Hence noninvertibility is a {\it witness} of non-Markovianity of the $\mathcal{C}$-class processes: operationally established non-invertibility implies non-Markovianity.
In this sense, for short, we say that invertibility is a
witness of Markovianity of the $\mathcal{C}$-class dynamical maps. Fortunately, invertibility of a dynamical map is straightforward operationally to test.
Concretely, as distinct from most of the existing Markovianity witnesses in the literature, testing invertibility does
not operationally require any kind of optimization.

Physical relevance of our considerations is at least three-fold. First, virtually all the basic physical laws are in the class $\mathcal{C}$. Second, equivalence of invertibility with time-locality of the  $\mathcal{C}$-class
dynamical maps provides a  classification of dynamical processes and {\it systematically} reproduces relations between the map invertibility, divisibility and time locality for the processes that have been investigated in the
literature--even beyond the topic of (non)Markovianity--while including the time-non-differentiable processes (as an example of the non-$\mathcal{C}$-class processes). Third,  the use of the witness of Markovianity for the $\mathcal{C}$-class processes is operationally a straightforward task.  Finally, a corollary of our
considerations is a derivation of the well-known Abel-Jacobi-Liouville identity \cite{arnoljd}.

In Section \hyperref[sec:2]{\ref{sec:2}} we introduce the $\mathcal{C}$ class of dynamical maps and prove the central result of this paper that is presented by Lemma 1 and illustrated by Figure 1.
In Section \hyperref[sec:2]{\ref{sec:3}} we place our conclusions in the context of
the various approaches to Markovianity as it can be found in the literature. To this end, as a support of our arguments, we provide certain technical details in Supplemental Material.
Section \hyperref[sec:2]{\ref{sec:4}} is discussion section
and we conclude in Section \hyperref[sec:2]{\ref{sec:5}}.

\section{Invertibility of the $\mathcal{C}$-class dynamical maps}\label{sec:2}

Basic physical laws are typically expected to be linear, continuous and smooth in time thus providing a differential mathematical form, i.e. a related differential equation whose solutions sufficiently describe dynamics and behavior of physical systems. These assumptions are so natural and therefore often only tacitly assumed. Nevertheless, those assumptions are our starting point that is formally introduced for the open-systems' dynamical maps.

\smallskip {\it Definition} 1. A {\it linear} and {\it completely positive}  dynamical map $\Phi$ is in the so-called $\mathcal{C}$ class of dynamical maps if and only if the following requirements are
simultaneously fulfilled: (a) the map is  time  continuous, in the sense it is  defined on a  continuous time interval $t'\in[t_{\circ},t]$, (b) the map is a two-parameter  map denoted $\Phi(t,t_{\circ}), t\ge t_{\circ}$,
(c) the map is smooth enough (ultraweak continuity), in the sense that,
for positive $\epsilon$, $\lim_{\epsilon\downarrow 0} \Phi(t+\epsilon,t_{\circ}) = \Phi(t,t_{\circ})$, for $t\ge t_{\circ}$, is well defined, (d) the map has the whole Banach space of statistical
operators (density matrices) in its domain,  and (e)  the map is differentiable, i.e. that the (ultraweak) limit:
\begin{equation}\label{eqn:2}
{d\Phi(t,t_{\circ})\over dt} =\lim_{\epsilon\downarrow 0} {\Phi(t+\epsilon,t_{\circ})-\Phi(t,t_{\circ})\over\epsilon} \tag{2.1}
\end{equation}

\noindent is well defined.

\noindent The items (a) and (c) are often assumed to be simultaneously satisfied. However, we want to emphasize existence of the discrete-time dynamical maps as well as the time-continuous processes
for which the map may not be well defined for some time instants.
The point (b) assumes the one-parameter families as special case, e.g. when the initial time instant $t_{\circ}$ is fixed or for the time-homogeneous processes when only the difference $t-t_{\circ}$ is of interest.
Every dynamical map that is not linear or non-completely positive or not satisfying at least some of the above conditions (a)-(e) of Definition 1 does not belong to the $\mathcal{C}$ class of dynamical maps.

According to Definition 1, the $\mathcal{C}$ class of dynamical maps is a subclass of the class of differentiable maps as defined solely by eq.(\ref{eqn:2}).
Differentiability introduced by eq.(\ref{eqn:2})  provides a differential equation, ${d\hat\rho(t)\over dt}={d\Phi(t,t_{\circ})\over dt}\hat\rho(t_{\circ})$, for the open system's state $\hat\rho(t)$,
where $\hat\rho(t_{\circ})$ is the initial state. Under the physically fairly general conditions,
equation (\ref{eqn:2}) allows for a master equation for the density matrix, $\hat\rho(t)$, of an open system to take the form \cite{breuer,rivas}:
\begin{equation}\label{eqn:3}
{d\hat\rho(t)\over dt} = -{\imath\over \hbar}[\hat H,\hat\rho(t)] + \int_{t_{\circ}}^t dr \mathcal{K}(t,r)\hat\rho(r) \tag{2.2}
\end{equation}

\noindent where $\hat H$ is the open-system's self-Hamiltonian, while the linear map $\mathcal{K}(t,r)$ is the ''memory kernel'' describing the effects of the environment on the system.

In eq.(\ref{eqn:3}), there is explicit dependence on the time instants $r<t$. However, Markovian processes are generally assumed {\it not} to carry any dependence on the time instants $r<t$ thus {\it requiring} a {\it time-local}  master equation of the general form of eq.(\ref{eqn:1}).

\subsection{A Markovianity witness}\label{sec:2.1}

The central result of this paper is the following lemma.

\smallskip {\it Lemma } 1. For a dynamical map $\Phi(t,t_{\circ})$ \textit{from the} $\mathcal{C}$ \textit{class} of dynamical maps, the following characteristics of the map are mutually {\it equivalent}:
(i) the map is invertible,  (ii) the map is divisible, and (iii) the map admits a time-local master equation.
\noindent

\smallskip We provide a proof of the lemma by establishing the chain of implications:
\begin{equation}\label{eqn:4}
(i) \Rightarrow (ii)\Rightarrow (iii) \Rightarrow (i).\tag{2.3}
\end{equation}

\noindent $(i)\Rightarrow (ii)$: Assuming existence of the inverse map, $\Phi^{-1}(t,t_{\circ})$, it easily follows
$\hat\rho(t)=\Phi(t,t_{\circ})\hat\rho(t_{\circ})=\Phi(t,t_{\circ})\Phi^{-1}(s,t_{\circ})\hat\rho(s), t\ge s\ge t_{\circ}$.
This expression presents a state  transition, $\hat\rho(s)\to\hat\rho(t)$, and introduces the map $\Phi(t,s)$ for this transition.
The requirement that everything regards arbitrary initial state $\hat\rho(t_{\circ})$ implies  divisibility of the map: $\Phi(t,s)=\Phi(t,t_{\circ})\Phi^{-1}(s,t_{\circ})$;

\noindent $(ii)\Rightarrow (iii)$: Assuming divisibility of the map,  equation (\ref{eqn:2})  leads (ultraweak continuity) to:
\begin{equation}\label{eqn:5}
{d\hat\rho(t)\over dt} = \lim_{\epsilon\downarrow 0}{\Phi(t+\epsilon,t_{\circ})-\Phi(t,t_{\circ})\over\epsilon}\hat\rho(t_{\circ})=
\lim_{\epsilon\downarrow 0}{\Phi(t+\epsilon,t)-\mathcal{I}\over\epsilon}\hat\rho(t):=
\mathcal{L}_t\hat\rho(t),\tag{2.4}
\end{equation}

\noindent which is a time-local master equation describing dynamics generated by the time-local Liouvillian $\mathcal{L}_t:= \lim_{\epsilon\downarrow 0}{\Phi(t+\epsilon,t)-\mathcal{I}\over\epsilon}$;

\noindent $(iii)\Rightarrow (i)$: The map $\Phi(t,t_{\circ})$ for which eq.(\ref{eqn:3}) applies can be presented due to the so-called time-splitting formula \cite{rivas,yosida} in the form:
\begin{equation}\label{eqn:6}
\Phi(t,t_{\circ}) = \lim_{\max\vert t'_{j+1}-t'_j \vert\to 0} \Pi_{j=n-1}^0 e^{\mathcal{L}_{t'_j}(t'_{j+1}-t'_j)},\tag{2.5}
\end{equation}

\noindent where $t=t'_n\ge t'_{n-1}\ge\dots\ge t_{\circ}$ and $\mathcal{L}_t$ is the Liouvillian in eq.(\ref{eqn:3}). Then the inverse is  constructed:
\begin{equation}\label{eqn:7}
\Phi^{-1}(t,t_{\circ}) = \lim_{\max\vert t'_{j+1}-t'_j \vert\to 0} \Pi_{0}^{j=n-1} e^{-\mathcal{L}_{t'_j}(t'_{j+1}-t'_j)},\tag{2.6}
\end{equation}

\noindent as it can be easily seen by inspection.

\noindent Not all requirements (a)-(e) in Definition 1 are equally relevant to all implications in the proof of Lemma 1. Taken together, those implications regard exactly the $\mathcal{C}$ class of dynamical maps.
This subtle point will be emphasized in Section \hyperref[sec:2]{\ref{sec:3.2}}, where we will recognize the Markovianity conditions  that can be found in the literature for the maps that are outside of
the $\mathcal{C}$ class of dynamical maps.

It is worth stressing,  that:    (A)  the product in equation (\ref{eqn:6}) (as well as in equation (\ref{eqn:7})) assumes the time-ordering thus providing the solution to equation (\ref{eqn:5})
in the standard exponential form, $\Phi(t,t_{\circ})=\mathcal{T}\exp\left(\int_{t_{\circ}}^t\mathcal{L}(s)ds\right)$, which is formally often used even for the continuous-variable systems \cite{breuer, rivas},
(B) the proof of Lemma 1 is {\it exact}, i.e. it may not apply  for certain approximation methods (e.g. perturbative approximation of Liouvillian) or short-time behavior (while bearing
in mind that for sufficiently short time-intervals, all dynamical maps are (approximately) invertible) \cite{rivas,breuer}, and (C) Lemma 1 implies the so-called Abel-Jacobi-Liouville identity \cite{arnoljd}
as demonstrated in Appendix \hyperref[sec:AppA]{\ref{sec:AppA}}.

Lemma 1 establishes  invertibility  as a {\it witness} of Markovianity: as distinguished above, non-invertibility, equivalently, non-divisibility, implies time non-locality and therefore non-Markovian
character for the $\mathcal{C}$-class dynamical maps. Nevertheless, just like equation (\ref{eqn:1}), Lemma 1 does not establish a sufficient condition for Markovianity of the $\mathcal{C}$-class
dynamical maps. This conclusion directly applies to concatenation of dynamical maps, $\Phi(t,t_{\circ})=\mathcal{V}(t,t_n)\circ\cdots\circ\mathcal{V}(t_2,t_1)\circ\mathcal{V}(t_1,t_{\circ})$.
Invertibility of every concatenated map $\mathcal{V}(t_j,t_i)$ in the class $\mathcal{C}$ is a necessary, but in general not sufficient condition for the total map $\Phi(t,t_{\circ})$ to be Markovian.

For the $\mathcal{C}$-class processes, complete state tomography \cite{nielsen} {\it suffices} for  determination of non-Markovian character of the process. That is, determining the final $\hat\rho(t)$
state for the given initial $\hat\rho(0)$ state  gives rise to the so-called ''process matrix'', $A$, defined (in a matrix representation) via $\rho(t) = A\rho(0)$. Invertibility of the process
is isomorphic with invertibility of the process matrix. Non-invertibility of the process matrix guarantees non-Markovian character of the $\mathcal{C}$ class dynamical maps. Hence, in principle, no
optimization procedure is required. To this end, as an illustration of the {\it  general} procedure, in Supplemental Material S1, we consider the well known one-qubit amplitude damping process \cite{nielsen}. For the
processes not belonging to the $\mathcal{C}$ class, Lemma 1 does not provide a proper procedure for determining non-Markovian character of the process.

\subsection{Comments}\label{sec:2.2}

As repeatedly emphasized, virtually all the basic physical laws are in the $\mathcal{C}$ class of dynamical maps, notably Newton's second law, the Hamilton's and Lagrange's equations of classical mechanics, the Maxwell
equations of classical electrodynamics as well as the time-dependent Schr\" odinger equation. That is, the basic physical laws are expected to be continuous in time as well as differentiable while free of any singularities
for every finite time instant $t'\in[t_{\circ},t]$. In this context, the dynamical models dealing with the discrete time instants should be regarded as approximations or the time-coarse-grained versions of the generic
physical processes.

Lemma 1 tells that the dynamical maps that are non-invertible but  local in time  fall out of the $\mathcal{C}$ class of dynamical maps. There is a sharp line dividing  the $\mathcal{C}$-class, and the
non-$\mathcal{C}$-class dynamical processes. Of all the invertible processes in the $\mathcal{C}$ class, only some of them may be Markovian. Those statements are illustrated by Figure 1.

\begin{figure*}[!ht]\label{fig:16.1}
\centering
    \includegraphics[width=0.4\textwidth]{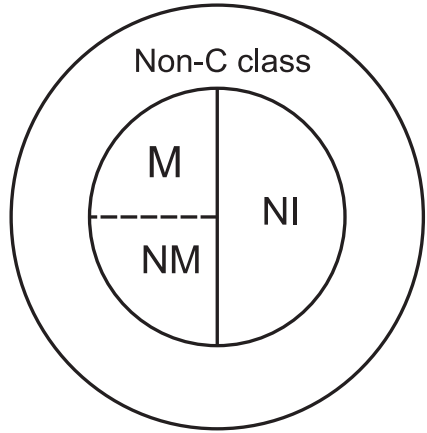}
\caption{A schematic presentation of the dynamical maps with the internal circle containing all the $\mathcal{C}$-class dynamical maps sharply divided from the non-$\mathcal{C}$-class dynamical maps (that are
out of the internal circle). The vertical solid line sharply divides invertible (the left part) from non-invertible (time-nonlocal, i.e. indivisible, the right part) $\mathcal{C}$-class processes. The dashed horizontal line sharply divides Markovian
(upper part) from the non-Markovian--invertible $\mathcal{C}$-class--processes. Position of the dashed line is not yet  uniquely determined--it depends on the adopted definition (criterion) of Markovianity. The used abbreviations are as follows: ''Non-C class'' stands for the non-$\mathcal{C}$-class processes; ''NI'' stands for ''noninvertible''
$\mathcal{C}$-class processes; ''M'' is for  ''Markovian'' while ''NM'' is for ''non-Markovian''  (invertible) $\mathcal{C}$-class processes.}
\end{figure*}

Invertibility as a witness of Markovianity of the $\mathcal{C}$-class of dynamical maps is easy operationally to use. The point is that no optimization is needed. In order to reduce the unavoidable error in
the quantum state tomography procedure, it may be useful to repeat the procedure or to combine it with the quantum process tomography \cite{nielsen,metrol}. Again, no optimization is required.

\section{Quantum Markovianity in context}\label{sec:3}

Figure 1 gives a {\it general framework} for classification of the dynamical maps.
The part ''NM'' in Figure 1 regards the $\mathcal{C}$-class invertible (and therefore time local) processes that are not Markovian. The ''NI'' part distinguishes
the $\mathcal{C}$-class noninvertible processes for which time-locality, i.e. eq.(\ref{eqn:1}), does not apply and therefore (according to eq.(\ref{eqn:1})) such processes {\it a priori} are non-Markovian. The ''Non-$\mathcal{C}$ class'' of dynamical maps contains
all the processes
that are out of the $\mathcal{C}$-class processes. Consequently, inapplicability of Lema 1 allows, in principle, various relations between time-locality, divisibility, invertibility and Markovianity of the maps.
This class of dynamical maps encompasses all the processes to which even eq.(\ref{eqn:3}) may not apply.

For the time local processes presented by eq.(\ref{eqn:1}), canonical form of the (time local) Liouvillian $\mathcal{L}_t$ is known  \cite{hall}:
\begin{equation}\label{eqn:8}
\mathcal{L}_t\hat\rho(t) = -{\imath\over\hbar}[\hat H,\hat\rho(t)]
+\sum_k \gamma_k(t) \left(\hat L_k(t)\hat\rho(t)\hat L_k^{\dag}(t) -{1\over 2}\{\hat L_k^{\dag}(t)\hat L_k(t),\hat\rho(t)\} \right),\tag{3.1}
\end{equation}

\noindent with Hermitian $\hat H$ and possibly time-independent damping factors $\gamma_k$ and the so-called Lindblad operators $\hat L_k$ that constitute an orthonormal set of traceless operators; the curly bracket
stands for the ''anticommutator''.

Bearing in mind Figure 1, it is now easy to recognize the characteristics of some of the time-local dynamical maps presented by eq.(\ref{eqn:8}). For example, if $\gamma_k(t)\ge 0, \forall{k}$ and for every
finite time instant $t'\in[t_{\circ},t]$, the map is said to be Markovian due to the RHP criterion \cite{rivas,rivasprl,rivas1}; in the special case of time independent $\gamma$s and time independent Lindblad operators,
eq.(\ref{eqn:8}) describes the celebrated semigroup of dynamical maps \cite{gorini}--the part ''M'' in Figure 1. If the condition $\gamma_k(t)\ge 0$ is not fulfilled for at least one index $k$ and one time instant $t'$,
the map is regarded non-Markovian due to the RHP criterion, thus belonging to the ''NM'' part in Figure 1. If for some finite time instant $t_{\ast}\in[t_{\circ},t]$ at least one damping factor $\gamma_k$
exhibits singularity (divergence), the map is out of the $\mathcal{C}$
class and necessarily non-invertible \cite{anders}--and should be placed in the part ''Non-$\mathcal{C}$ class'' in Figure 1.

Hence we can say that  all the dynamical maps presented in the literature can be easily recognized in Figure 1. To this end, the only remaining question is the {\it criterion} of Markovianity, i.e.
the ''position'' of the dashed line in Figure 1--above,
we used the RHP criterion, which is only one out of a numerous set of such criteria \cite{rivas1,breuer1,spanci,wiseman,milz}. Therefore, bearing in mind that Lemma 1 provides a necessary, but not a sufficient condition for Markovianity of dynamical maps--due to the criterion eq.(\ref{eqn:1}), we know for sure what is {\it not} Markovian--we proceed by analyzing the known results in the light of the sharp line (internal circle in Figure 1) dividing
the $\mathcal{C}$-class from the non-$\mathcal{C}$-class
dynamical maps.

There is  a large number of possibilities for breaking the assumptions of Definition 1 and hence the statement of Lemma 1. It is practically impossible to present all of them in a single paper.
For this reason we proceed by placing some prominent theoretical models and results in the context of Lemma 1.
In Section \hyperref[sec:2]{\ref{sec:3.1}} we consider
the role of the $\mathcal{C}$-class processes for the CP-divisibility and BLP conditions of quantum Markovianity. In Section \hyperref[sec:2]{\ref{sec:3.2}} we focus on determining
which assumptions of Definition 1 are not fulfilled for certain processes that are regarded Markovian--that includes the continuous but non-invertible processes describable by
a time-local master equation. Thence a clear role of the particular assumptions of Definition 1 in the proof of Lemma 1.

\subsection{The $\mathcal{C}$-class processes}\label{sec:3.1}

Lemma 1 establishes divisibility as a necessary condition for Markovianity of a $\mathcal{C}$-class dynamical map. Since the RHP criterion (CP-divisibility) requires the map divisibility,
it naturally appears as a proper criterion for Markovianity of the $\mathcal{C}$-class processes. Then a Markovian map $\Phi(t,t_{\circ})$ is defined by the requirement of complete positivity
of the intermediate map $\Phi(t,s), s>t_{\circ}$, thus necessarily being representable by eq.(\ref{eqn:8}) with the non-negative damping factors $\gamma_k(t')$ for every instant of time
$t'\in[t,t_{\circ}]$ \cite{rivas,rivasprl,kosak}. In Supplemental Material S1  we provide an illustration of the general procedure for determining CP-divisibility for the case of the one-qubit amplitude-damping process.

Divisibility as a necessary condition for Markovianity  places some restrictions on usefulness of the BLP criterion of Markovianity for the $\mathcal{C}$ class of dynamical processes.
Actually, necessity of divisibility of a map effectively reduces the BLP to the RHP criterion of Markovianity. This is, the only remaining distinction between the BLP and the RHP
criteria is the possibility that the intermediate map (which always exists for the $\mathcal{C}$-class processes)  may not be completely positive; then (as emphasized above)
at least some of the damping factors take negative values for certain time instants (or intervals) \cite{rivas,rivasprl,anders,kosak}. However, this possibility does not seem in keeping with intuition for Markovian
processes \cite{milz,rivas,utagi1,gardiner,kosak1,alicki,shaji} (especially in analogy with the classical concept of Markovianity \cite{rivas1,milz,utagi1,kosak1}),
and certainly cannot lead to equation (\ref{eqn:8}). Example processes from the $\mathcal{C}$ class are known \cite{hall,utagi1}, for which the BLP criterion fails to
detect Markovianity established by equation (\ref{eqn:8}).

In the more general context, some divisible but non-CP-divisible maps may carry some (possibly weak) memory \cite{rivas1,breuer1,spanci,wiseman,milz,rivas,breuer,utagi1,gardiner,kosak1,alicki,kosak2}
of the previous history thus presenting non-Markovian processes. This kind of memory disappears for the time-homogeneous Markovian processes (time-independent terms
in equation (\ref{eqn:8})), which is thus recognized as a stronger concept of memorylesness \cite{utagi1} than   CP-divisibility \cite{rivas,rivasprl,rivas1}.

Invertibility as a necessary condition for Markovianity is, in a sense, at  variance with some information-theoretic considerations of Markovianity \cite{datta}.
In this context it is argued \cite{datta} [our emphasis]: ''As such, it is clear that the assumption of bijectivity [{\it i.e. map invertibility}] eludes a purely information theoretic or operational
description and must be put 'by hand' on top of the dynamical evolution.''
According to Lemma 1, for the $\mathcal{C}$ class of dynamical maps  this is \textit{not} the case. Furthermore,
for the $\mathcal{C}$-class processes, operational test of non-Markovianity reduces to the standard state-tomography procedures (Section \hyperref[sec:2]{\ref{sec:2}}). Therefore it seems that  a search for Markovianity
beyond invertibility (e.g. for certain information-processing tasks) should regard processes out of the $\mathcal{C}$ class of dynamical maps. Thence it is no surprise that going beyond invertibility
often targets the time-discrete processes (cf. Section \hyperref[sec:2]{\ref{sec:3.2}}), in which context the BLP criterion obtains a natural generalization \cite{datta}, thus possibly suggesting that the BLP criterion
for Markovianity
is well suited for certain processes out of the $\mathcal{C}$ class of dynamical maps.

\subsection{Beyond the $\mathcal{C}$-class processes}\label{sec:3.2}

%In this section we analyze consequences of non-validity of the items (a)-(e) in Definition 1, thus going beyond the $\mathcal{C}$ class of dynamical maps and, consequently, {\it beyond applicability of} Lemma 1.In Figure 1, this is indicated by ''Non-$\mathcal{C}$ class''. For this class of dynamical maps, various relations between time-locality, divisibility, invertibility and Markovianity of theprocess may be allowed--e.g. there may appear noninvertible processes that admit a time-local dynamics as per eq.(\ref{eqn:1}), and also there may be some maps not being describable by any kind of master equation.Nevertheless, as presented by the implication (i)$\Rightarrow$(ii) in the proof of Lemma 1, invertibility implies divisibility of the map.

There are not any universal relations between time locality, divisibility and invertibility for the processes out of the $\mathcal{C}$ class of dynamical maps. Expectedly, there is not a general definition or the operational procedures (witnesses) for determining Markovianity.

%We perform by considering non-validity of those items separately--i.e. assume that at least one of the requirements ((a)-(e) in Definition 1) for the $\mathcal{C}$-class dynamical maps is not fulfilled.

In order to demonstrate this statement, we separately analyse non-validity of the items (a)-(e) in Definition 1, thus going beyond the $\mathcal{C}$ class of dynamical maps.
In support of the general notes (distinguished by ''(GN)'')
made below, we consider only a small portion of the relevant literature while providing some technical details in Supplemental Material.

\subsubsection{Continuity}

\noindent (GN) Dynamical maps which are not time-continuous do not allow for introducing the time derivative eq.(\ref{eqn:1}) and therefore the master equation formalism does not apply. Consequently,
eq.(\ref{eqn:5}) does not apply and therefore  the implications (ii)$\Rightarrow$(iii) and (iii)$\Rightarrow$(i) in the proof of Lemma 1 fail. Below we distinguish some scenarios for non-continuous dynamical evolutions.

\noindent {\it Continuity 1}. The maps obtained by combining continuous ''quantum channels'' from the $\mathcal{C}$ class can fall outside of the $\mathcal{C}$ class of dynamical maps and therefore
the Lindblad-kind of Markovianity (i.e. CP-divisibility) for the mixed channels is lost. In general, this applies to both the convex mixing of channels as well as to the subsequent operations of
the channels. For example, mixing the CP-divisible Pauli channels (one-qubit channels) gives rise to a CP-indivisible channel \cite{jaga,jaga1,wudarski}.
In Supplemenatl Material S2  we provide a single-qubit model that illustrates this argument.

\noindent {\it Continuity 2}.  Every discrete map as well as certain concatenations of (possibly continuous) dynamical maps break the assumption of the time-continuous dynamics. Those include the possibility
of external (e.g. due to experimenters actions) interruptions by resetting the open system's states, or by performing quantum measurement in the course of the system's evolution as well as endowing the continuous
dynamics by occasional classical stochastic events (often termed ''quantum jumps''), which effects in a non-deterministic time evolution. Then the quantum-maps Markovianity should be investigated independently of Lemma 1.
In Ref. \cite{datta}, the authors identify divisibility with ''information-theoretic Markovianity'' without invoking invertibility of the considered dynamical maps. Nevertheless, as long as a continuous-time
approximation is allowed for the considered discrete dynamical maps, this kind of Markovianity should be taken with caution: the continuous-time limit may be subject of Lemma 1. That is, albeit divisibility
is regarded a key feature of Markovianity, in the continuous-time limit it may become equivalent with invertibility and therefore (cf. Section \hyperref[sec:2]{\ref{sec:3.1}}) resorting to CP-divisibility as a formal definition of Markovianity.
On the other hand, in the operational approach that introduces Markovianity as a statement about multitime correlations (thus introducing multi-time dynamical maps) \cite{pollock1,pollock2,modi}, certain
interventions on the open system are allowed.
Those external interventions on the open system {\it not} involving any time-continuous dynamics extends the standard framework (Section \hyperref[sec:2]{\ref{sec:3.1}}), from which perspective it is found that  CP-divisibility
does not necessarily guarantee the absence of memory \cite{modi}. Exclusion of  continuous dynamics makes this approach and the concept of Markovianity complementary to the  one distinguished by Lemma 1.

\subsubsection {Two-parameter family}

\noindent (GN) Introducing additional parameters for a dynamical map gives rise to introducing a whole family of families (of subfamilies) of the two-parameter maps. Additional parameter (or parameters),
denoted $\alpha$, that can be time dependent, imposes the task of analyzing applicability of  Lemma 1 to every such subfamily, $\Phi_{\alpha}(t,t_{\circ})$, separately. In general, different values of the
parameter $\alpha$ may lead to different characteristics of the two-parameter (sub)families regarding invertibility, divisibility and time-locality of the related master equations (if such exist).
That is, validity of the implications in the proof of Lemma 1 should be separately investigated for every subfamily. In effect, in this regard, the total family $\{\Phi_{\alpha}(t,t_{\circ})\}$ of dynamical
maps may not be described by the definite statements.

This is characteristic e.g. of the so-called ''tensor power'' of dynamical maps \cite{filipov1} as well as for ''time deformations'' of master equations \cite{filipov2}, where an additional
parameter $\alpha$ counts the evolution families $\Phi_{\alpha}(t,0)$. Those maps (subfamilies of the two-parameter maps) taken separately are subject of Lemma 1 and can exhibit either CP or non-CP,
depending on $\alpha$. Taken together, those maps cannot be uniquely determined as CP or non-CP, or time-local or time-non-local. Hence Markovianity of the maps cannot be decided either.
Analogous findings apply for the three-parameter maps introduced and
investigated in \cite{jaga,jaga1}. In Supplemental Material S2  we distinguish an example of a three-parameter family of dynamical maps.

\subsubsection{Smoothness}

\noindent (GN) If, for some time instants, or intervals, the map is not sufficiently smooth,  the derivative eq.(\ref{eqn:1}) is not well defined. Consequently, eq.(\ref{eqn:5}) does not apply
in the vicinity of such time instants/intervals thus implying inapplicability of the implication (ii)$\Rightarrow$(iii) in the proof of Lemma 1. The presence of singularities implies the map is  noninvertible \cite{anders}.
However, in general, it is not implied that a time-local generator cannot  be defined for such processes. It is only implied that such generators (Liouvillians) {\it must},if they exist, carry singularity. For such cases, existence
of a time-local master equation and the map non-invertibility is not a priori excluded.
In this regard it is a paradigmatic example of a two-level system (e.g. an atom) decay in the {\it finite} time intervals. Intuitively, the physical picture is appealing: up to some finite time instant $t_{\ast}$,
the time-continuous dynamics is regular and fulfills all the assumptions of Definition 1, i.e. Lemma 1 applies. For the time instants $t\ge t_{\ast}$, the system does not evolve in time as it is already in the ground
state and the map is simply the identity map. In effect, the map is neither everywhere differentiable nor invertible, but is (obviously) divisible. To this end, illustrative examples can be found in Supplemental Material S3.

The models of time-local master equation with a singularity-carrying Liovillians are known \cite{breuer,hrusc1,hrusc,hall1}. Validity of eq.(\ref{eqn:2}) opens in principle the door for Markovianity of certain
such processes. In Ref. \cite{hrusc}, the authors regard extension of CP-divisibility for the non-smooth dynamical maps. They find  that giving up the condition of invertibility may provide validity of equation (\ref{eqn:8})
for the {\it price} of non-positivity of some damping factors $\gamma_i$ in equation (\ref{eqn:8}), at least for some time instants/intervals. Analogous results, yet starting from investigating invertibility of the map,
can be found in Ref. \cite{anders}.  At this instance, the two criteria for Markovianity, RHP \cite{rivas1,rivas} and the BLP \cite{breuer1,backflow} criterion, are recognized mutually equivalent \cite{hrusc}
for non-invertible  dynamical maps, which are outside of the $\mathcal{C}$ class of dynamical maps.

\subsubsection {The map domain}

\noindent (GN) If a map is with restricted domain in the Banach space of statistical operators, the implication (i)$\Rightarrow$(ii) of the proof of Lemma 1 does not apply. Thence invertibility is not necessarily equivalent
with divisibility of the map. For such dynamical maps, even the condition of complete positivity (i.e. existence of the Kraus integral form of the process), which is a basic requirement for Markovianity \cite{breuer,rivas},
may be at stake.

Reducing the map domain from the whole Banach space to a subspace is closely linked with the role of the initial tensor-product state for equation (\ref{eqn:8}) to follow in the microscopic derivations \cite{rivas,breuer}.
This subtle topic \cite{sabani,broduc} reveals close connection of the initial correlations in the ''system+environment'' isolated system with complete positivity of the open $S$-system's dynamics. For restricted
domain of certain dynamical maps, even the presence of initial correlations may lead to completely positive dynamics with the map non-CP for the states out of the map's domain. If restriction to the reduced domain may be
allowed then Figure 1 can be directly applied.

\subsubsection {Differentiability}

\noindent (GN) Nonexistence of the map derivative introduced by eq.(\ref{eqn:2}) implies inapplicability of eq.(\ref{eqn:5}) and hence of the implications (ii)$\Rightarrow$(iii) and (iii)$\Rightarrow$(i) in the proof of Lemma 1.
Inapplicability of eq.(\ref{eqn:5}) may imply nonexistence of the map inverse as well as nonexistence of a proper master equation for the process--as already emphasized by the above item ''1'' (''continuity''). As a particularly
interesting case that applies for the continuous and smooth dynamical evolution, we emphasize the maps for which the necessary condition for differentiability, $\Phi(t_{\circ},t_{\circ})\neq \mathcal{I}$,
is not fulfilled.

For a specific kind of non-differentiable and noninvertible but divisible  maps \cite{PRSA1,PRSA2} for which $\Phi(t_{\circ},t_{\circ})\neq \mathcal{I}$,  the standard CP-divisibility  can be dynamically established without invoking the
condition of invertibility. This comes at the price of impossibility to introduce a differential form, i.e. a master equation, for the  map, despite the fact that it is a unital map preserving the identity operator.
Dynamical emergence of CP-divisibility regards the long time intervals, in contrast to the standard CP-divisibility of Section \hyperref[sec:2]{\ref{sec:3.1}}. Therefore Markovianity understood as CP-divisibility is an emergent and conditional
characteristic of the open system dynamics.

\section{Discussion}\label{sec:4}

Equation (\ref{eqn:1}) is a broad criterion for Markovianity for all differentiable processes. That is, for the differentiable processes we know for sure which dynamical maps are non-Markovian.

The following additions to the general debate on quantum Markovianity are presented in this paper: (a) the class of dynamical
processes for which Markovianity requires invertibility of the process (Section \hyperref[sec:2]{\ref{sec:2}}),
(b) questioning usefulness of the BLP criterion for Markovianity regarding the $\mathcal{C}$ class dynamical maps (Section \hyperref[sec:3.1]{\ref{sec:3.1}}), and (c) a fresh view on
the concept of Markovianity beyond   invertibility (i.e. beyond the $\mathcal{C}$ class of dynamical maps, Section \hyperref[sec:2]{\ref{sec:3.2}}).

The standard (non)Markovianity witnesses  typically assume optimization procedures over the set of states or measurement-operators where the quantum state tomography is often the first step \cite{rivas1,breuer1,spanci,wiseman,milz}.
The witness of non-Markovianity introduced in this paper reduces to the state tomography without any further steps or procedures, including the optimization. To this end, it suffices to know in advance that the
process is in the $\mathcal{C}$ class of dynamical evolution and to obtain non-invertibility in order to detect non-Markovianity of the process; as repeatedly emphasized in this paper, observation of invertibility
does not guarantee Markovianity of the process. Operationally, managing the errors inherent in the state tomography procedures as well as in the counting statistics can be performed in the same vein as in the standard
witnesses of non-Markovianity, e.g. as in the experiments \cite{rivas1,breuer1} employing the ''trace distance'' (the BLP criterion) to witness non-Markovianity.

It is essential to re-emphasize: Lemma 1 is {\it exact}. That is, certain approximations of the map itself, or of the Liouville superoperator, or considerations of short (or not-very-short) time intervals may,
in principle, vary the conclusions that follow from the exact treatment (some perturbation methods are known  to may jeopardize complete positivity of the map \cite{breuer,rivas}). On the one hand, an approximation may not capture certain relevant characteristics of the exact map. On the other hand, different approximation
techniques are used to arrive at ''Markovian'' master equations, and most of them are implicit to Section \hyperref[sec:2]{\ref{sec:3.2}}: everything depends on the adopted meaning of ''Markovian dynamics'' \cite{wiseman}. While more detailed analysis
of alternative definitions of Markovianity on the basis of Lemma 1 is certainly interesting, extension of the basic remarks made in Section \hyperref[sec:2]{\ref{sec:3}} is far beyond the present paper and certainly cannot be properly presented
in a single paper. For this reason we do not discuss the Markovianity witnesses that are widely used in the literature.

Finally, we want to stress, that the class $\mathcal{C}$ of dynamical maps as simply and clearly defined by Definition 1 may appear a rather subtle concept in applications: it need not be obvious if a map
is in the $\mathcal{C}$ class. On the one hand, [as emphasized in Section \hyperref[sec:2]{\ref{sec:3.2}}], some discrete maps admitting the  continuous-time limit may fall within this class.
On the other hand, mixing of the dynamical maps outside of the $\mathcal{C}$ class can  give a map effectively in the class $\mathcal{C}$ \cite{wudarski,utagi}. Interestingly, it is known \cite{utagi} that mixing of
some singularity-carrying continuous dynamical maps may wipe the singularities out and generate a map falling within the $\mathcal{C}$ class of dynamical maps. It is out of doubt that those examples are not exceptional
and some other kinds of effectively $\mathcal{C}$-class dynamical maps may be expected to be found or recognized in the near future. All of them will be subject of Lemma 1 as long as the possibly additional assumptions
or constraints on the map are in accord with Definition 1.

\section{Conclusion}\label{sec:5}

Mutual relations of divisibility, invertibility, time-locality and Markovianity of the quantum dynamical maps cannot be presented in simple terms.
Nevertheless, we present a systematic approach by introducing and analyzing a physically relevant class of dynamical maps for which
invertibility is a necessary condition for Markovianity. Operationally, in principle, no optimization is required for determining non-Markovian character of the distinguished-class of quantum processes.
On this basis it is now possible to more closely analyze the origin of the existing criteria and formal definitions of quantum Markovianity, as they can be found in the literature.

\noindent {\bf Acknowledgements}
The present work was supported by The Ministry of Education, Science and Technological Development of
the Republic of Serbia (451-03-68/2022-14/ 200122) and in part for MD by the ICTP-SEENET-MTP project NT-03 Cosmology--Classical and Quantum Challenges.

\appendix

\section{A derivation of the Abel-Jacobi-Liouville identity}\label{sec:AppA}

Lemma 1 implies for a dynamical map $\Phi(t,t_{\circ})=\Phi(t,s) \Phi(s,t_{\circ})$, where $\Phi(t,s)=\Phi(t,t_{\circ}) \Phi^{-1}(s,t_{\circ})$,
while $\Phi(t,t_{\circ})=\mathcal{T} e^{\int_{t_{\circ}}^t\mathcal{L}(u)du}$.
Then, with the use of the standard relations for the matrix determinants, an isomorphic matrix representation of the dynamical maps gives:
$\det\Phi(t,t_{\circ}) = \det\Phi(t,t_{\circ})\det\Phi^{-1}(s,t_{\circ})$  $\det\Phi(s,t_{\circ})$.
On the use of the time-splitting formula and the equality $\det e^{A}=e^{tr A}$ easily follows:
\begin{equation}\label{eqn:A1}
\det\Phi(t,t_{\circ}) = \det\Phi(s,t_{\circ}) e^{\int_s^t tr\mathcal{L}(u)du}, \tag{A.1}
\end{equation}

\noindent which is the Abel-Jacobi-Liouville identity (often presented for $t_{\circ}=0$) \cite{arnoljd}.

The inverse does not, in general, hold. That is, equation (\ref{eqn:A1}) admits both, $\Phi(t,t_{\circ})=\Phi(t,s) \Phi(s,t_{\circ})$ and $\Phi(t,t_{\circ})= \Phi(s,t_{\circ})\Phi(t,s)$ that,
in general, is not correct--it is correct e.g. for the semigroup dynamical maps. Similarly, from equation (\ref{eqn:A1}) it may seem allowed to write $\Phi(t,s)=\mathcal{T} e^{\int_s^t\mathcal{L}(u)du}$,
which also, in general, is not correct. Therefore equation (\ref{eqn:A1}) does not imply Lemma 1.

\pagebreak

\pagebreak

\noindent {\bf Figure captions} A schematic presentation of the dynamical maps with the internal circle containing all the $\mathcal{C}$-class dynamical maps sharply divided from the non-$\mathcal{C}$-class dynamical maps (that are
out of the internal circle). The vertical solid line sharply divides invertible (the left part) from non-invertible (time-nonlocal, i.e. indivisible, the right part) $\mathcal{C}$-class processes. The dashed horizontal line sharply divides Markovian
(upper part) from the non-Markovian--invertible $\mathcal{C}$-class--processes. Position of the dashed line is not yet  uniquely determined--it depends on the adopted definition (criterion) of Markovianity. The used abbreviations are as follows: ''Non-C class'' stands for the non-$\mathcal{C}$-class processes; ''NI'' stands for ''noninvertible''
$\mathcal{C}$-class processes; ''M'' is for  ''Markovian'' while ''NM'' is for ''non-Markovian''  (invertible) $\mathcal{C}$-class processes.

\end{document}

\centerline{\bf Supplemental Material for: }
\centerline{\bf Invertibility as a witness of Markovianity }
\centerline{\bf of the quantum dynamical maps}

\bigskip

\centerline{  Jasmina Jekni\' c-Dugi\' c$^a$, Momir Arsenijevi\' c$^b$, Miroljub Dugi\' c$^{\ast b}$}

\bigskip

$^a$University of Ni\v s, Faculty of Science and Mathematics, Vi\v
segradska 33, 18000 Ni\v s, Serbia

$^b$University of Kragujevac, Faculty of Science, Radoja
Domanovi\' ca 12, 34000 Kragujevac, Serbia

\section{One-qubit amplitude damping process}\label{sec:S1}

Consider the amplitude damping process, which is in the class $\mathcal{C}$ of dynamical maps for a single qubit defined by the following Kraus operators \cite{nielsen}:
\begin{equation}
E_{\circ}= \begin{pmatrix}
1 & 0 \\
0 & e^{-\gamma t}
\end{pmatrix}, \quad
E_1= \begin{pmatrix}
0 & \sqrt{1-e^{-2\gamma t}} \\\nonumber
0 & 0
\end{pmatrix},\tag{S1.1}
\end{equation}

\noindent that are given in the Pauli $z$-representation, $Z=\begin{pmatrix}
-1 & 0 \\
0 & 1
\end{pmatrix}$.

For the arbitrary initial qubit state given in the same representation, $\rho(0)={1\over 2} \begin{pmatrix}
1-n_z & n_+ \\
n_- & 1+n_z
\end{pmatrix}$, the final state in an instant of time $t$  reads:
\begin{equation}\nonumber
\rho(t)= E_{\circ}\rho(0) E_{\circ} + E_1\rho(0) E_1^{\dag} =
{1\over 2} \begin{pmatrix}
2-(1+n_z)e^{-2\gamma t} & n_+e^{-\gamma t} \\
n_-e^{-\gamma t} & (1+n_z)e^{-2\gamma t}
\end{pmatrix}
\end{equation}

Isomorphically presented as the matrix-columns, the initial and the final state give rise to the process matrix $A$, defined as $A\rho(0)=\rho(t)$, that can be easy found to read:
\begin{equation}
A= \begin{pmatrix}
1 & 0 & 0 & 1-e^{-2\gamma t}\\
0 &  e^{-\gamma t} & 0 & 0\\
0 & 0 & e^{-\gamma t} & 0\\
0 & 0 & 0 & e^{-2\gamma t}
\end{pmatrix}\tag{S1.2}
\end{equation}

\noindent  Eigenvalues of the matrix $A$, $1, e^{-\gamma t}, e^{-2\gamma t}$, are all nonzero (except in the asymptotic limit of $t\to\infty$). Therefore the matrix is non-singular, i.e. the process represented by the matrix $A$ is invertible.

Complete positivity of the process requires positivity of the related, so-called, dynamical matrix $B$. The general recipe for obtaining the dynamical matrix from the process matrix, $A\to B$, reads \cite{sudar}:
\begin{equation}
A=\begin{pmatrix}
p_1 & p_2 & p_3 & p_4\\
P_1 & P_2 & P_3 & P_4\\
q_1 & q_2 & q_3 & q_4\\
Q_1 & Q_2 & Q_3 & Q_4
\end{pmatrix} \to
B=\begin{pmatrix}
p_1 & p_2 & P_1 & P_2\\
p_3 & p_4 & P_3 & P_4\\
q_1 & q_2 & Q_1 & Q_2\\
q_3 & q_4 & Q_3 & Q_4
\end{pmatrix}\tag{S1.3}
\end{equation}

\noindent Applying equation (S1.3) to (S1.2) gives the dynamical matrix:
\begin{equation}
B= \begin{pmatrix}
1 & 0 & 0 & e^{-\gamma t}\\
0 & 1-e^{-2\gamma t} & 0 & 0\\
0 & 0 & 0 & 0\\
e^{-\gamma t} & 0 & 0 & e^{-2\gamma t}
\end{pmatrix}\tag{S1.4}
\end{equation}

\noindent whose eigenvalues, $0,1\pm e^{-2\gamma t}$, are non-negative. Therefore the matrix $B$ is positive and hence the process is CP--as we already know from the very existence of the Kraus form for the process.

From equation (S1.2) follows the inverse matrix:
\begin{equation}
A^{-1}= \begin{pmatrix}
1 & 0 & 0 & 1-e^{2\gamma t}\\
0 &  e^{\gamma t} & 0 & 0\\
0 & 0 & e^{\gamma t} & 0\\
0 & 0 & 0 & e^{2\gamma t}
\end{pmatrix}\tag{S1.5}
\end{equation}

\noindent that now gives rise to the matrix representation of the process for a nonzero initial instant of time $s$:
\begin{equation}\nonumber
A(t,s) = A(t) A^{-1}(s) = \begin{pmatrix}
1 & 0 & 0 & 1-e^{-2\gamma (t-s)}\\
0 &  e^{-\gamma (t-s)} & 0 & 0\\
0 & 0 & e^{-\gamma (t-s)} & 0\\
0 & 0 & 0 & e^{-2\gamma (t-s)}
\end{pmatrix},
\end{equation}

\noindent which defines the dynamical matrix:
\begin{equation}
B(t,s)= \begin{pmatrix}
1 & 0 & 0 & e^{-\gamma (t-s)}\\
0 & 1-e^{-2\gamma (t-s)} & 0 & 0\\
0 & 0 & 0 & 0\\
e^{-\gamma (t-s)} & 0 & 0 & e^{-2\gamma (t-s)}
\end{pmatrix}\tag{S1.6}
\end{equation}

\noindent whose eigenvalues, $0, 1\pm e^{-2\gamma(t-s)}$, are all non-negative thus
establishing complete positivity for the process starting in a non-zero instant of time $s$.
Existence of $A(t,s)$ and the fact that it's CP means that the amplitude damping process is CP-divisible.

\section{A convex combination of the $\mathcal{C}$-class dynamical maps}\label{sec:S2}

Consider a convex combination of the $\mathcal{C}$ class processes for one qubit \cite{jaga},
$\Phi_i(t)[\hat\rho(0)]\equiv \Phi_i(t, 0)[\hat\rho(0)]:= (1-p(t))\hat\rho(0) + p(t)\hat\sigma_i \hat\rho(0)\hat\sigma_i, i=y,z$:
\begin{equation}
\Phi(t, 0) = a \Phi_z(t, 0) + (1-a) \Phi_y(t, 0),\tag{S2.1}
\end{equation}

\noindent where $p(t)=(1-e^{-rt})/2\in[0,1/2)$ and $\hat\sigma_i, i=y,z$ represent the Pauli sigma-operators and the probability $0< a< 1$.

Then the total process reads:
\begin{equation}\nonumber
\Phi(t, 0)[\hat\rho(0)] = (1-p(t))\hat\rho(0) +ap(t)\hat\sigma_z\hat\rho(0)\hat\sigma_z + (1-a)p(t)\hat\sigma_y\hat\rho(0)\hat\sigma_y.
\end{equation}

On the use of the procedure described in Section \hyperref[sec:S1]{\ref{sec:S1}} follows the process matrix:
\begin{equation}
A(t,0)=\begin{pmatrix}
1-(1-a)p & 0 & 0 & (1-a)p\\
0 & 1-(1+a)p & (a-1)p & 0\\
0 & (a-1)p & 1-(1+a)p & 0\\
(1-a)p & 0 & 0 & 1-(1-a)p
\end{pmatrix}\tag{S2.2}
\end{equation}

\noindent whose eigenvalues are nonzero: $1,1-2p,1-2ap,1-2p+2ap$. Therefore the process is invertible. The easy obtained dynamical map for equation (S2.2) has the following, non-negative eigenvalues: $0,2(1-p), 2ap, 2p(1-a)$. Therefore the process (as it is known in advance) is CP.

For the process starting in a non-zero instant of time $s$, the matrix $A(t,s)$ can be straightforwardly obtained while giving rise to the dynamical matrix of the process:
\begin{equation}
B(t,s)= {1\over 2}\begin{pmatrix}
1+\alpha & 0 & 0 & \beta+\gamma\\
0 & 1-\alpha & \beta-\gamma & 0\\
0 & \beta-\gamma & 1-\alpha & 0\\
\beta+\gamma & 0 & 0 & 1+\alpha
\end{pmatrix}\tag{S2.3}
\end{equation}

\noindent where: $\alpha=(2ap-2p+1)/(2aq-2q+1), \beta=(1-2p)/(1-2q), \gamma=(1-2ap)/(1-2aq)$, while $p=1-e^{-rt} > q =1-e^{-rs}$ for $t>s$.

From equation (S2.2) it is obvious, that the map $\Phi$ is a {\it three}-parameter process (for the initial zero instant of time), $\Phi_{r,a}(t)$, thus not belonging to the $\mathcal{C}$ class of dynamical maps.
For the dynamical matrix $B(t,s)$ the following eigenvaules can be obtained: $b_1=1+\alpha-\beta-\gamma, b_2=1-\alpha+\beta-\gamma, b_3=1-\alpha-\beta+\gamma$ and $b_4=1+\alpha+\beta+\gamma$.
Thence the condition of positivity of the matrix $B(t,s)$, i.e. of the complete positivity of the process reads: $\vert 1\pm\alpha \vert \ge \vert \beta\pm\gamma\vert$. For the parameter values not fulfilling the constraint,
the matrix $B$ is non-positive and hence the $A(t,s)$ is non-CP. Therefore this three-parameter process is neither CP-divisible nor non-CP-divisible.
In accordance with Lemma 1, for the fixed values of the parameter $a$, the map becomes a two-parameter process and therefore exhibits either CP-divisibility, or non-CP-divisibility--depending on
the choice of values of the parameters.

\section{Examples of time non-smooth  processes}\label{sec:S3}

\noindent Example 1. Consider the process defined by the matrix elements of a one-qubit statistical operator $\hat\rho(t)$ \cite{hrusc1}:
\begin{equation}\label{eqn:S31}
\begin{split}
\rho_{00}(t)=\rho_{00}(0)x_{\circ}(t) + \rho_{11}(0)(1-x_1(t))\\
\rho_{11}(t)=\rho_{00}(0)(1-x_{\circ}(t))+ \rho_{11}(0)x_1(t)\\
\rho_{01}(t)=\rho_{01}(0)\gamma(t)
\end{split}\tag{S3.1}
\end{equation}

Performing the procedure described in Section \hyperref[sec:S1]{\ref{sec:S1}} follows the process matrix $\mathcal{A}$ as well as the related dynamical matrix $\mathcal{B}$ (for simplicity, we place: $x_i\equiv x_i(t)$):
\begin{equation}\label{eqn:S32}
\mathcal{A}=\begin{pmatrix}
x_{\circ} & 0 & 0 & 1-x_1\\
0 & \gamma & 0 & 0\\
0 & 0 & \gamma^{\ast} & 0\\
1-x_{\circ} & 0 & 0 & x_1
\end{pmatrix}, \quad \mathcal{B}=\begin{pmatrix}
x_{\circ} & 0 & 0 & \gamma\\
0 & 1-x_1 & 0 & 0\\
0 & 0 & 1-x_{\circ}& 0\\
\gamma^{\ast} & 0 & 0 & x_1
\end{pmatrix}.\tag{S3.2}
\end{equation}

Eigenvalues of the $\mathcal{B}$ matrix read:  $1-x_i, i=0,1$ and $(x_{\circ}+x_1\pm\sqrt{4\gamma^2+(x_{\circ}-x_1)^2})/2$. In order the process be completely positive, non-negative eigenvalues of $\mathcal{B}$ appear under the conditions: $x_i\in[0,1], i=0,1$, and $\vert\gamma\vert^2\le x_{\circ}x_1$. On the other hand, eigenvalues of the process matrix: $1,\gamma,x_{\circ}+x_1-1$. Therefore, for the time instants for which $\gamma=0$ and/or $x_{\circ}+x_1=1$, the process matrix $\mathcal{A}$ is singular and hence the process is non-invertible. This conclusion also follows from the form of the Liouvillian (we use the original notation of Ref. \cite{hrusc1}):
\begin{equation}\label{eqn:S33}
\mathcal{L}[\hat\rho] = -\imath{\Omega\over 2}[\hat\sigma_z,\hat\rho]+\sum_{k=+,-}a_k\mathcal{L}_k[\hat\rho]+{\Gamma\over 2}\mathcal{L}_z[\hat\rho],\tag{S3.3}
\end{equation}

\noindent where appear the Pauli sigma-operators, $\hat\sigma_i$, as formally the Lindblad operators  for the ''Liuouvillians'' $\mathcal{L}_i, i=+,-,z$, and
\begin{equation}\label{eqn:S34}
\begin{split}
\Gamma:=- {a_{\circ}+a_1\over 2}-Re{\dot\gamma\over\gamma}, \quad \Omega:=Im{\dot\gamma\over\gamma}\\
a_{\circ}:={\dot x_{\circ}(1-x_1)+\dot x_1x_{\circ}\over 1-x_{\circ}-x_1}, \quad a_1:={\dot x_1(1-x_{\circ})+\dot x_{\circ}x_1\over 1-x_{\circ}-x_1}\end{split}
\tag{S3.4}
\end{equation}

\noindent That is, for the time instants for which the $\mathcal{A}$ matrix is singular, the functions $\Gamma(t)$ and $\Omega(t)$ diverge thus implying
\textit{singularities} for the Liouvillian, i.e. non-smooth dynamics in the vicinity of those time instants. Needless to say, if the functions $x_{\circ}$ and $x_1$
are such that $\gamma(t)\neq 0 \neq 1-x_{\circ}(t)-x_1(t), \forall{t}$, the process is invertible and the master equation well defined by the Liouvillian equation
(\ref{eqn:S33}) of the Lindblad form (i.e. of the Markovian, CP-divisible, form of equation eq.(3.1))--in accordance with Lemma 1.

\noindent Example 2. Consider the map defined by the  following matrix representation of the one-qubit state \cite{breuer,hrusc}:
\begin{equation}\label{eqn:S35}
\rho(t)=\begin{pmatrix}
\vert G(t)\vert^2\rho_{11}(0) & G(t)\rho_{12}(0)\\
G^{\ast}(t)\rho_{21}(0) & (1-\vert G(t)\vert^2)\rho_{11}(0)+\rho_{22}(0)
\end{pmatrix}\tag{S3.5}
\end{equation}

The process matrix $\mathcal{A}$ and the dynamical matrix $\mathcal{B}$ read (we put $G\equiv G(t)$):
\begin{equation}\label{eqn:S36}
\mathcal{A}=\begin{pmatrix}
\vert G\vert^2 & 0 & 0 & 0\\
0 & G & 0 & 0\\
0 & 0 & G^{\ast} & 0\\
1-\vert G(t)\vert^2 & 0 & 0 & 1
\end{pmatrix}, \quad \mathcal{B}=\begin{pmatrix}
\vert G\vert^2 & 0 & 0 & G\\
0 & 0 & 0 & 0\\
0 & 0 & 1-\vert G(t)\vert^2 & 0\\
 G^{\ast} & 0 & 0 & 1
\end{pmatrix}.\tag{S3.6}
\end{equation}

\noindent Eigenvalues of the dynamical matrix read: $0,1\pm \vert G\vert^2$. Therefore we conclude that the process will be completely positive if and only if $0\le \vert G(t)\vert^2\le 1, \forall{t}$.
The eigenvalues of the process matrix, $1,G,G^{\ast},\vert G\vert^2$, can be nonzero, and the process invertible, only if $G(t)\neq 0,\forall{t}$. Supposing \cite{hrusc} that there exists a finite
time instant $t_{\ast}$ such that $G(t)\neq 0, t<t_{\ast}$, while $G(t)=0, t\ge t_{\ast}$, the process matrix is singular and hence the (total) process is non-invertible. In analogy with the
previous example, the Lindblad-form Liouvillian for the process carries singularity for $t=t_{\ast}$ \cite{hrusc}:
\begin{equation}\label{eqn:S37}
\mathcal{L}[\hat\rho] = -\imath{s(t)\over 2}[\hat\sigma_+\hat\sigma_-,\hat\rho] + \gamma(t)\left(\hat\sigma_-\hat\rho\hat\sigma_+ -{1\over 2}\{\hat\sigma_+\hat\sigma_-,\hat\rho\}  \right),\tag{S3.7}
\end{equation}

\noindent where: $s(t) =-2Im{\dot G\over G}$ and $\gamma(t)=-2Re{\dot G\over G}$. Interestingly, the process (\ref{eqn:S35}), albeit non-invertible, is divisible:
$\Phi(t,0)=\Phi(t,t_{\ast})\Phi(t_{\ast}-\epsilon,0)$ for $\epsilon\to 0^+$, where $\Phi(t,t_{\ast})=\mathcal{I}, t\ge t_{\ast}$. Needless to say, for $G(t)\neq 0, \forall{t}$,
the map is invertible, $\dot G(t)\le 0$ and the Liouvillian equation (\ref{eqn:S37}) is well defined for the completely positive process whose master equation is of the form of
equation (1.1)--in accordance with Lemma 1.